\documentclass[11pt,twoside]{article}
\usepackage{asp2010}
\resetcounters

\def\etal{{\it et al.~}}
\def\eg{{\it e.g.,~}}

\def\kms{~{\rm km~s^{-1}}}
\def\cm3{~{\rm cm^{-3}}}

\begin{document}
\title{Cosmic Ray Spectrum in Supernova Remnant Shocks}

\author{Hyesung~Kang
\affil{Department of Earth Sciences, Pusan National University, Pusan 609-735,
Korea}
}

\begin{abstract} 

We performed kinetic simulations of diffusive shock acceleration
in Type Ia supernova remnants (SNRs) expanding into a uniform 
interstellar medium (ISM).
The preshock gas temperature is the primary parameter that governs 
the cosmic ray (CR) acceleration,  
while magnetic field strength and CR injection rate are secondary parameters. 
SNRs in the hot ISM, with an injection fraction smaller than $10^{-4}$, 
are inefficient accelerators with less than 10 \% energy getting converted 
to CRs. The shock structure is almost test-particle like and the 
ensuing CR spectrum can be steeper than $E^{-2}$. 
Although the particles can be accelerated to the knee energy of $10^{15.5}Z$eV
with amplified magnetic fields in the precursor, 
Alfv\'enic drift of scattering centers softens the source spectrum as steep 
as $E^{-2.1}$ and reduces the CR acceleration efficiency.

\end{abstract}

\section{Introduction} 

Most of Galactic cosmic rays up to at least the knee energy of $10^{15.5}$eV, 
are believed to be accelerated at SNRs within our Galaxy 
by diffusive shock acceleration (DSA) \citep[see][and references therein]{hill05}. 
In DSA theory, a small fraction of incoming thermal particles can be injected 
into the CR population, and accelerated to very high energies through their 
interactions with resonantly scattering Alfv\'en waves
in the converging flows across the shock \citep{md01}. 
Kinetic simulations of the CR acceleration at SNRs
have shown that up to 50 \% of explosion energy can be converted to CRs,
when a fraction $10^{-4}-10^{-3}$ of incoming thermal particles are injected
into the CR population at the gas subshock \citep{bv97,kang06}.
This should be enough to replenish the galactic CRs escaping from our Galaxy
with $L_{CR}\sim 10^{41} {\rm erg ~s^{-1}}$.

Multi-band observations of nonthermal radio to $\gamma$-ray
emissions from several SNRs have been successfully explained by 
efficient DSA features such as high degree of shock compression 
and the possible magnetic field amplification in the precursor 
\citep{reynolds08,berezhko09,morlino09}.
High-resolution X-ray observations of several young SNRs exhibit very thin rims,
indicating the presence of magnetic fields as strong as a few $100 \mu$G 
downstream of the shock \citep{pariz06}.
Moreover, theoretical studies have shown that efficient magnetic field amplification
via resonant and non-resonant wave-particle interactions is an integral part of DSA
\citep{lb00, bell04}.
In the downstream region magnetic fields can be amplified by turbulence that is induced
through cascade of vorticity generated behind curved shocks \citep{giacal07}.
If there exist such amplified magnetic fields in the upstream region of SNRs,
CR ions with change $Z$ might gain energies up to $E_{\rm max} \sim 10^{15.5}Z$ eV,
which may explain the all-particle CR spectrum up to the second knee
at $\sim10^{17}$ eV with rigidity-dependent energy cutoffs.

It has been recognized, however, that the CR spectrum at sources, 
$N(E)\propto E^{-s}$ with $s<2$, predicted based on the nonlinear DSA may 
be too flat to be consistent with the observed flux of CR nuclei at Earth, 
$J(E) \propto E^{-2.7}$.
Assuming an energy-dependent propagation path length ($\Lambda \propto E^{-0.6}$),
a softer source spectrum with $s \sim 2.3-2.4$, is preferred by the observed 
data \citep[\eg][]{ave09}.
This discrepancy could be reconciled if one consider 
the drift of scattering centers with respect to the bulk plasma \citep{ski75}.
It reduces the velocity jump that the particles experience across
the shock, which in turn softens the CR spectrum beyond the canonical 
test-particle slope \citep{kang10, capri10, pzs10}.

Using the spherical CRASH (Cosmic-Ray Amr SHock) code, 
we have calculated the CR spectrum accelerated at SNRs 
from Type Ia SNe expanding into a uniform interstellar medium.

\section{Numerical Calculation}

\subsection{Spherical CRASH Code in an Expanding Grid}

In the kinetic equation approach to numerical study of DSA,
the following diffusion-convection equation for the particle momentum
distribution, $f(r,p,t)$, is solved along with suitably modified gasdynamic
equations 
\begin{equation}
{\partial g\over \partial t}  + (u+u_w) {\partial g \over \partial r}
= {1\over{3r^2}} {\partial \over \partial r} \left[r^2 (u+u_w)\right]
\left( {\partial g\over \partial y} -4g \right) 
+ {1 \over r^2}{\partial \over \partial r} \left[r^2 \kappa(r,y)
{\partial g \over \partial r}\right],
\label{diffcon}
\end{equation}
where $g=p^4f$, with $f(p,r,t)$ the pitch angle averaged CR
distribution, $u_w$ is the wave speed,
and $y=\ln(p)$, and $\kappa(r,y)$ is the diffusion coefficient
parallel to the field lines \citep{ski75}.

The basic gasdynamic equations and details of the spherical CRASH 
code can be found in \citet{kj06}. 
The advection term of Eq. (\ref{diffcon}) is solved by the wave-propagation 
method, as for the gasdynamic variables, except that only the entropy
wave applies. Then the diffusion term is solved by the semi-implicit
Crank-Nicholson scheme. 
In order to implement the shock tracking and AMR (Adoptive Mesh Refinement)
techniques effectively in a spherical geometry, we solve the fluid and 
diffusion-convection
equations in a comoving frame that expands with the outer shock.
Since the shock is at rest and tracked accurately as a true discontinuity, 
we can refine the region around the gas subshock at an arbitrarily fine level.
Moreover, the shock remains at the same location in the comoving grid,
so the compression rate is applied consistently to the CR distribution
at the subshock. 
This results in much more accurate and efficient low energy CR acceleration
and faster numerical convergence on coarser grid spacings,
compared to the simulations in a fixed, Eulerian grid.

\begin{table}[!ht]
\caption{Model Parameters for Supernova Remnants}
\begin{center}
{\small
\begin{tabular}{ lrrrrrrrr }
\tableline
\noalign{\smallskip}
Model & $n_H$ & $T_0$ & ~$B_{\mu}$  & $r_o$ &~~$t_o$ &
  $u_o$ &$P_o$  \\
~ & ($\cm3$) & (K) & ($\mu$G)  & (pc) & (years) &
($10^4 \kms$) & ($10^{-6}$erg cm$^{-3}$) \\
\noalign{\smallskip}
\tableline
\noalign{\smallskip}
WISM  & 0.3   & $3.3\times 10^4$ & 30 & 3.19 & 255. &1.22 & 1.05 \\
MISM  & 0.03  &$10^5$            & 30 & 6.87 & 549. &1.22 & $1.05\times10^{-1}$ \\
HISM  & 0.003 &$10^6$           & 5 & 14.8 & 1182. &1.22 & $1.05\times10^{-2}$ \\
\noalign{\smallskip}
\tableline
\end{tabular}
}
\end{center}
\end{table}

\begin{figure}[!ht]
\vskip -0.5cm
\plotone{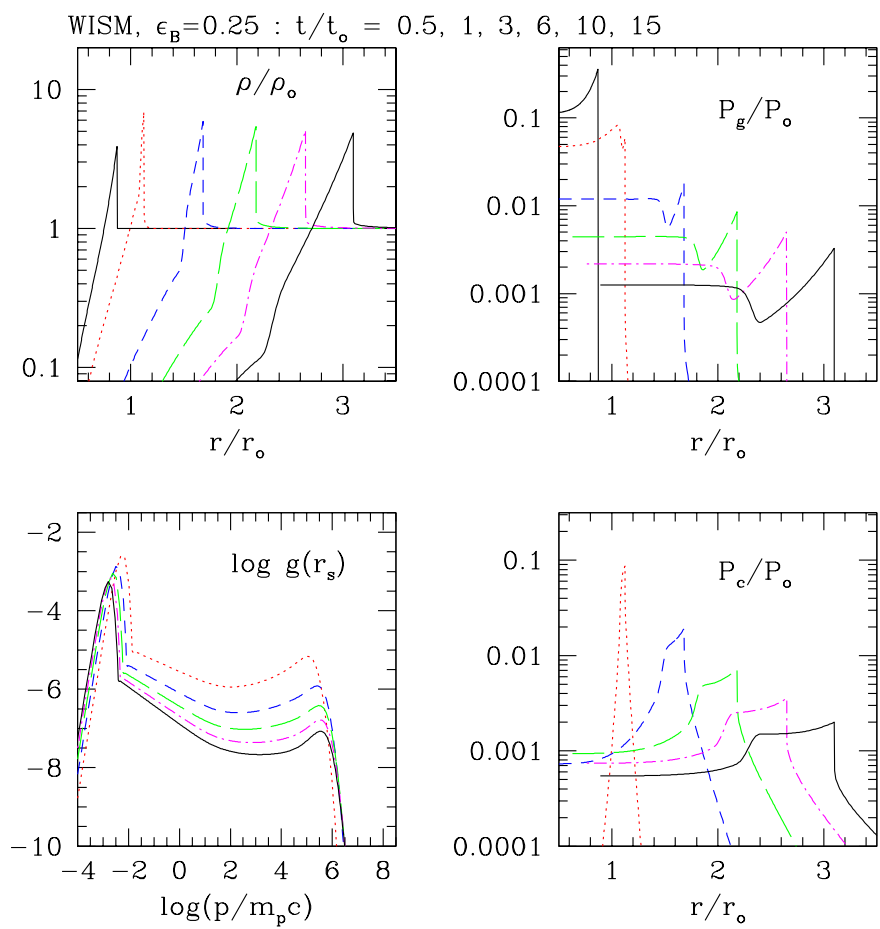}
\vskip -0.5cm
\caption{
Time evolution of the SNR model in the warm phase ISM.
The CR distribution function at the subshock, $g_s=f_s(p)p^4$, is shown as well.
}
\vskip -0.6cm
\end{figure}

\begin{figure}[!ht]
\vskip -0.5cm
\plotone{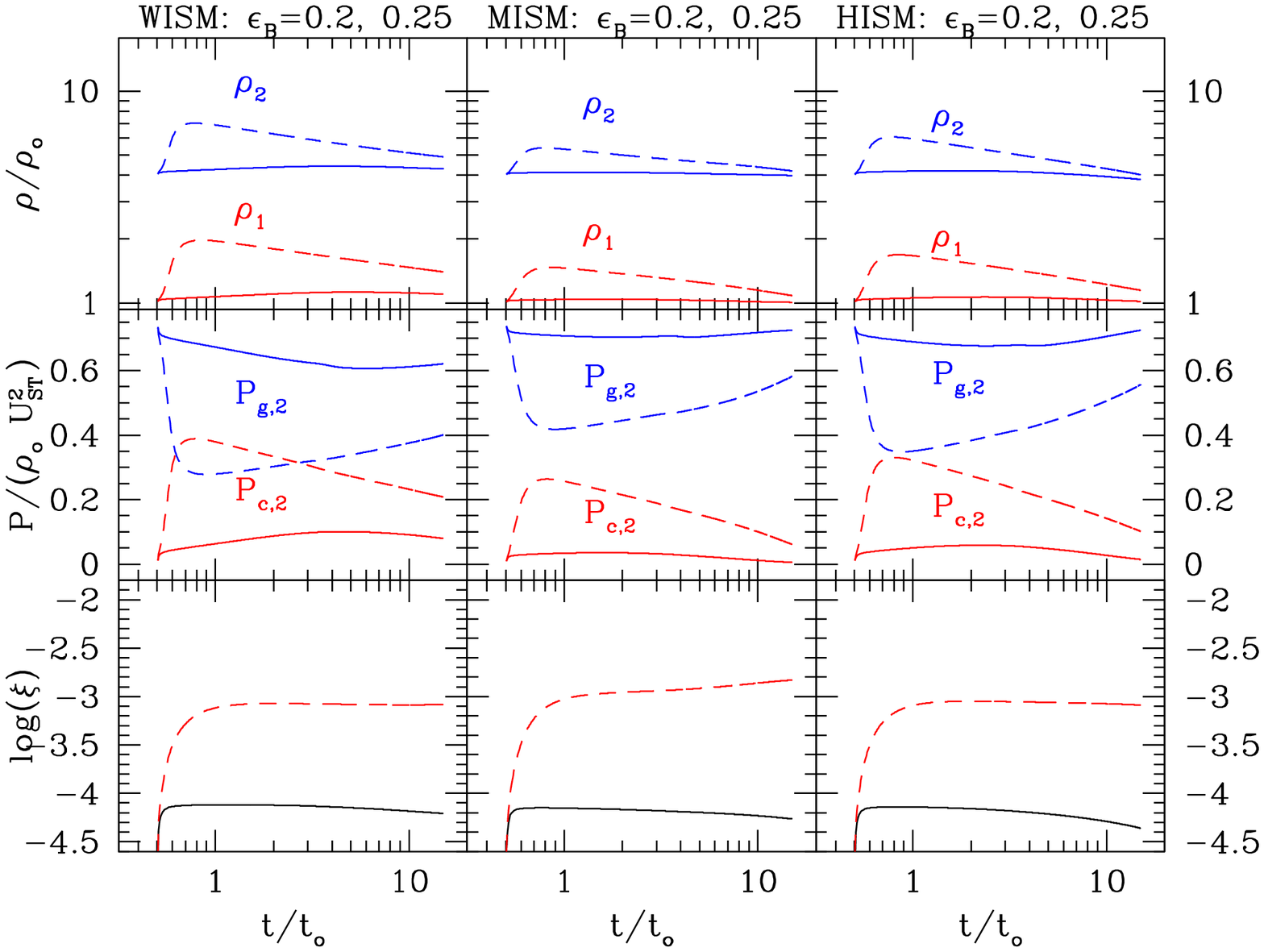}
\vskip -3.0cm
\caption{
Immediate pre-subshock density, $\rho_1$, post-subshock density, $\rho_2$,
post-subshock CR and gas pressure in units of the ram pressure of the
unmodified Sedov-Taylor solution, $\rho_0 U_{ST}^2$,
and the CR injection fraction, $\xi$
are plotted for models WISM (left), MISM (middle), and HISM (right).
Two values of $\epsilon_B=0.20$ (solid lines) and $\epsilon_B=0.25$ (dashed lines), are adopted.
}
\vskip -0.5cm
\end{figure}

At quasi-parallel shocks, small anisotropy in the particle
velocity distribution in the postshock fluid frame causes some particles
in the high energy tail of the Maxwellian distribution to stream upstream \citep{gbse92, md01}.
This thermal leakage injection process is treated numerically by adopting a 
{\it phenomenological} injection scheme, in which particles above a
certain injection momentum $p_{\rm inj}$ cross the subshock and get
injected to the CR population \citep{kjg02}.
One free parameter controls this process;
$\epsilon_B = B_0/B_{\perp}$, the ratio of
the general magnetic field along the shock normal, $B_0$, to
the amplitude of postshock MHD wave turbulence, $B_{\perp}$ \citep{mv98}.
This parameter controls the fraction of particles
injected into the CR population, $ \xi$, at the gas subshock.
On the other hand, \citet{giacalone05} showed that the protons can
be injected efficiently even at perpendicular shocks in fully
turbulent fields due to field line meandering.

Assuming that particles are resonantly scattered by self-generated waves,
we adopt a Bohm-like diffusion model that represent a saturated wave spectrum,
$\kappa(r,p) = \kappa_{\rm n}\cdot (p/{m_p c}) (\rho_0/\rho)$,
where $\kappa_n= m_p c^3/(3eB_0)= 3.13\times 10^{22} {\rm cm^2s^{-1}} (B_0/\mu{\rm G})^{-1}$.

\subsection{Alfv\'en Wave Transport}

The scattering by Alfv\'en waves tends to isotropize the CR distribution 
in the wave frame, which may drift upstream at the Alfv\'en speed, 
$u_w = v_A = B_0/ \sqrt{4\pi \rho}$, with respect to the bulk plasma flow,
where $B_0$ is the amplified magnetic field strength \citep{ski75}.
In the postshock region, $u_w = 0$ is assumed, since the Alfv\'enic
turbulence in that region is probably relatively balanced.
This Alfv\'enic drift reduces the velocity difference between 
upstream and downstream scattering centers compared to the bulk flow, 
leading to less efficient DSA.
So the `modified' test-particle slope can be estimated as
$q_{\rm tp} = 3(u_0 - v_A) / (u_0-v_A -u_2)$, 
where $u_2$ is the downstream flow speed \citep{md01}.
Hereafter we use the subscripts '0', '1', and '2' to denote
conditions far upstream, immediately upstream 
and downstream of the subshock, respectively.
Thus the drift of Alfv\'en waves in the upstream region tends to soften
the CR spectrum from the canonical test-particle spectrum of $f(p)\propto p^{-4}$,
if the Alfv\'en Mach number ($M_A= u_s/v_A$) is small.

\begin{figure}[!ht]
\vskip -0.5cm
\plotone{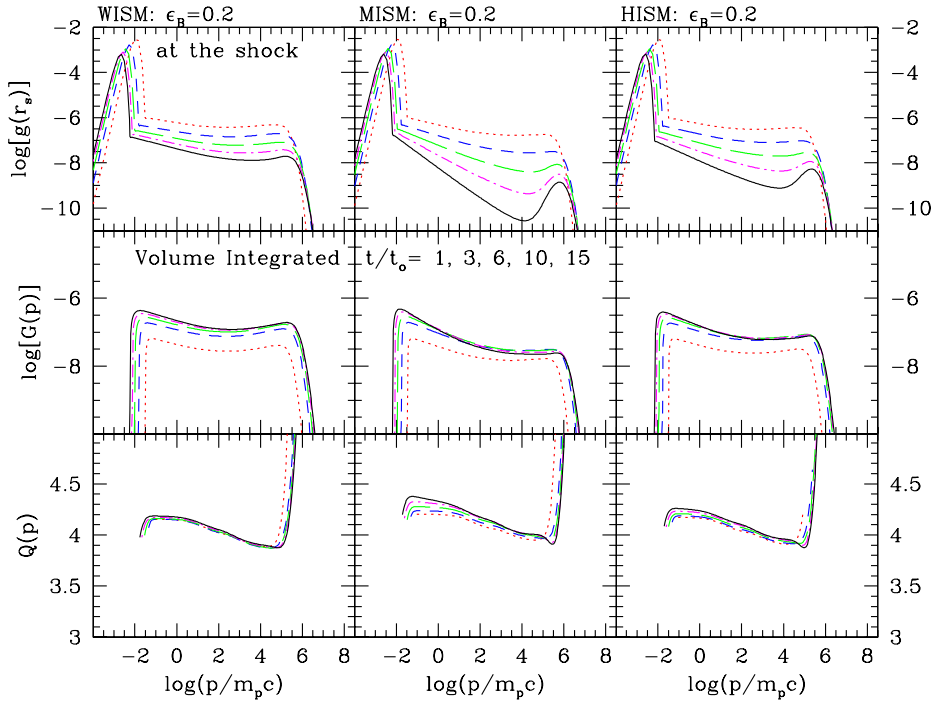}
\vskip -3.0cm
\caption{
The CR distribution at the shock, $g(r_s,p)$, the volume integrated CR number, $G(p)$, 
and its slope, $Q(p)$, are shown for models with $\epsilon_B=0.20$. 
}
\vskip -0.5cm
\end{figure}

In addition, gas heating due to Alfv\'en wave dissipation in the upstream region is
considered by the term $W(r,t)= -  v_A ({\partial P_c / \partial r})$
where $P_c$ is the CR pressure.
This term is derived from a simple model in which Alfv\'en waves are amplified by
streaming CRs and dissipated locally as heat in the precursor region.
This effect reduces the subshock Mach number thereby reducing DSA efficiency
\citep{bv97,kj06}. 

\subsection{Model Parameters for Type Ia Supernova Remnants}

We consider a Type Ia SN explosion with the ejecta mass,
$M_{ej}=1.4 M_{\sun}$, expanding into a uniform ISM.
All models have the explosion energy, $E_o=10^{51}$ ergs.
The shock sonic Mach number is the key parameter determining the evolution and the DSA efficiency,
while the particle injection rate and magnetic field strength (through $v_A$ and
$\kappa_n$) are secondary parameters.
So here three phases of the ISM are considered:
the {\it warm phase} with $T_0=3\times 10^4$K (WISM),
the {\it intermediate phase} with $T_0=10^5$K (MISM),
and the {\it hot phase} with $T_0=10^6$K (HISM).
The presence of amplified magnetic fields a few $\times 100 \mu$G
downstream of shock has been indicated by very thin rims of several young 
SNRs observed in X-ray \citep[\eg][]{reynolds08}. 
To represent this effect we take the upstream field strength, $B_0=30\mu$G
for the WISM and MISM models. 
For the HISM model, however, the SNR shock is much weaker and not dominated by the CR pressure, 
so the filed amplification should be minimal. Thus the mean ISM field of $5\mu$G is adopted.
The physical quantities are normalized by the following constants:
$r_o=\left({3M_{ej}/ 4\pi\rho_o}\right)^{1/3}$,
$t_o=\left({\rho_o r_o^5 / E_o}\right)^{1/2}$, 
$u_o=r_o/t_o$,
$\rho_o = (2.34\times 10^{-24} {\rm g cm^{-3}}) n_H$,
and $P_o=\rho_o u_o^2$.
Model parameters are summarized in Table 1.
Two values of $\epsilon_B=0.20$ and $\epsilon_B=0.25$ are adopted for
inefficient and efficient injection cases, respectively.

\begin{figure}[!ht]
\vskip -0.5cm
\plotone{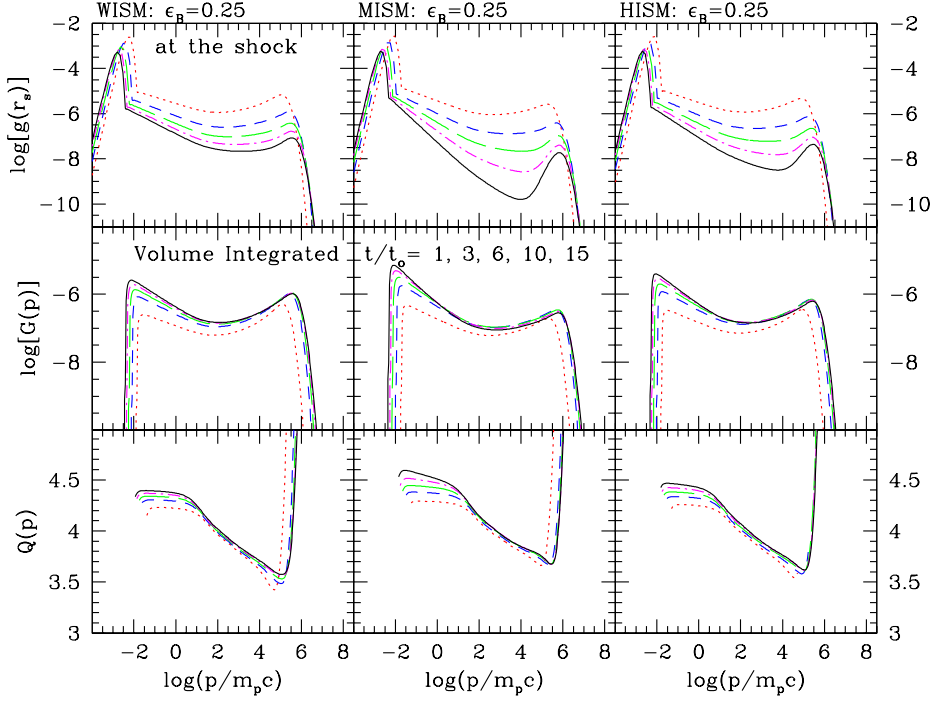}
\vskip -3.0cm
\caption{
Same as Figure 3 except $\epsilon_B=0.25$.
}
\vskip -0.5cm
\end{figure}

\section{Results}

Figure 1 shows the time evolution of the WISM model with $\epsilon_B=0.25$. 
The injection is very efficient with $\xi \approx 10^{-3}$, 
the shock becomes dominated by $P_c$ 
and the total density compression peaks at $\rho_2/\rho_0 \approx 6$
at $t/t_o \sim 1$. Afterward, the CR acceleration becomes less efficient
as the shock slows down and becomes weaker. 

Figure 2 shows the evolution of shock properties in the different models.
In the models with $\epsilon_B=0.2$ (solid lines), the CR injection and acceleration are inefficient with
the injected CR fraction, $\xi \approx 10^{-4}$, and the postshock CR pressure,
$P_{c,2}/(\rho_0 U_{\rm ST}^2) < 0.1$ through out the entire Sedov-Taylor (ST) stage.
In such inefficient acceleration regime, the shock remains test-particle like
with $\rho_2/\rho_0 \approx 4$ and $\rho_1/\rho_0\approx 1$.
In the models with $\epsilon_B=0.25$ (dashed lines), on the other hand, $\xi \approx 10^{-3}$, 
the ratio $P_{c,2}/(\rho_0 U_{\rm ST}^2)$ reaches up to 0.4, and 
the total density compression peaks at $\rho_2/\rho_0 \approx 7$. 
These two ratios decrease in time, since both the sonic and Alfv\'enic Mach numbers decrease.
The WISM model is the more efficient accelerator than the HISM model, because of the higher
sonic Mach number. 
Despite the lower sonic Mach number, in the HISM model 
the CR acceleration is more efficient, compared to the 
MISM model, because of weaker Alfv\'enic drift effects due to lower $B_0$.  

\begin{figure}[!ht]
\vskip -0.7cm
\plotone{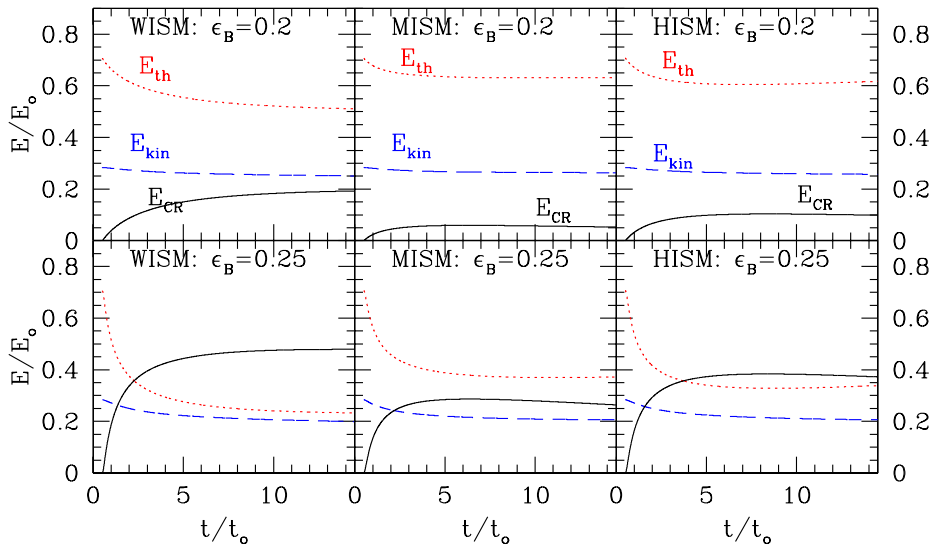}
\vskip -5.0cm
\caption{
Total thermal, kinetic and CR energies inside the simulation volume
in units of the explosion energy $E_o$ for the models shown in Figure 2.
}
\vskip -0.6cm
\end{figure}

Figures 3-4 show the CR distribution function at the shock, $g(r_s,p)$,
the volume integrated CR spectrum, $G(p) = \int 4\pi g(r,p) r^2 {\rm d}r$,
which represents the spectrum of the particles confined by the shock
and its slope $Q(q) = - d \ln G/ d \ln p + 4 $.
In all models the cutoff momentum $p_{\rm max}$ approaches up to 
$\sim 10^{15}-10^{15.5}$ eV/c. 
We note that, if the shock follows the Sedov-Taylor solution,
the cutoff momentum asymptotes to $p_{\rm max}/m_pc \approx 0.61(u_o^2 t_o/\kappa_n) 
\sim 10^{6.5}$ at large $t$, which corresponds to $E_{\rm max} \approx 10^{15.5}$ eV 
for the WISM model.
Even for the inefficient injection cases with $\xi \la 10^{-4}$ the CR spectrum 
at the shock exhibit concave curvatures as a consequence of
momentum dependent diffusion across the precursor
and slowing-down of the spherical shock.
Softening of the spectrum at low energies is further enhanced by 
the Alfv\'enic drift and weakening of the subshock strength in time,
which is more significant in the cases of stronger $B_0$.
However, the concavity of $G(p)$ is much less pronounced than that
of $g(r_s,p)$. 
In the inefficient injection models, 
the slope, $Q(p)$, varies 4.2-4.3 at low momentum  
and 3.9-4.2 at high momentum,  
and $G(p)$ does not changes significantly for $t/t_o \ga 6$,
especially for $10^{2}m_pc<p <p_{\rm max}$.
This implies that the acceleration is nearly balanced by the adiabatic cooling
during the late stage. 
So we can predict that the form of $G(p)$ would remain roughly the
same at much later time.
In the efficient injection models with $\xi \ga 10^{-3}$, of course,
nonlinear feedback effects are more substantial.

Figure 5 shows the integrated energies,
$E_i/E_o = 4\pi \int e_i(r) r^2 {\rm d}r$, where $e_{th}$,
$e_{kin}$, and $e_{CR}$ are the densities of thermal, kinetic and
cosmic ray energy, respectively.
The total CR energy accelerated by $t/t_o= 15$ is
$E_{\rm CR}/E_o=$ 0.15, 0.05, and 0.08 for WISM, MISM, and HISM models, 
respectively, for $\epsilon_B=0.2$.
These are marginally consistent with the requirement that 
an order of 10 \% of SN explosion energy needs to be
converted to CRs to replenish the Galactic CRs.
In the efficient injection models with $\epsilon_B=0.25$,
the CR energy fraction approaches to $E_{\rm CR}/E_o=$ 0.45, 0.25, and 0.35
for WISM, MISM, and HISM models, respectively.
The CR acceleration in the warm 
ISM model with $\epsilon_B=0.25$ may be somewhat too efficient.
But one has to recall that the CR injection rate may depend
on the mean magnetic field direction relative to the shock surface.
In a more realistic magnetic field geometry, where a uniform
ISM field is swept by the spherical shock, only 10-20 \% of the
shock surface has a quasi-parallel field geometry \citep{volk03}.
Moreover, Type Ia SNe make up only about 15 \% of SN explosions in the Galaxy,
while core collapse SNe exploding inside wind bubbles may behave differently
\citep{hill05}.

\section{Conclusion} 

In general, the DSA is very efficient for strong SNR shocks, if
the injection fraction is $\xi \ga 10^{-3.5}$.
The CR spectrum at the subshock shows a strong concavity, not only
because the shock structure is modified nonlinearly by the dominant
CR pressure, but also because the spherical SNR shock slows down in time during the
Sedov-Taylor stage.
Thus the concavity of the CR spectrum in SNRs is more
pronounced than that in plane-parallel shocks.
Moreover, the Alfv\'enic drift in the precursor further softens
the CR spectrum as the Alfv\'enic Mach number decreases. 
However, the volume integrated spectrum, $G(p)$ (the spectrum
of CRs confined by the shock including the particles in the upstream
region) is much less concave.
In the test-particle like solutions,
$G(p)$ approaches roughly to time-asymptotic states in the late Sedov-Taylor stage,
since the acceleration is nearly balanced by adiabatic cooling.

If the injection fraction is $\xi \ga 10^{-3}$,
about 25-45\% of the explosion energy is transferred to
CRs and the source CR spectrum becomes $N(E)\propto E^{-s}$ with 
$s \approx 1.6-1.8$ for $10^{11}< E< 10^{15}$ eV, which might be
too flat to be consistent with the observed CR spectrum at Earth \citep{ave09}.
If $\xi \la 10^{-4}$, on the other hand,
the shock structure is almost test-particle like with $\rho_2/\rho_0 \approx 4.2-4.4$
and the predicted source spectrum has a slope $s \approx 2.0-2.1$.
However, the fraction of CR energy conversion, $E_{\rm CR}/E_0 \approx 0.05-0.15$,
might be only marginally consistent with the Galactic CR luminosity. 

Finally, in all models considered in this study, for Bohm-like diffusion 
with the amplified magnetic field in the precursor, 
the particles could be accelerated to $E_{\rm max} \approx 10^{15.5}Z$eV.
The drift and dissipation of {\it faster} Alfv\'en waves in the precursor, 
on the other hand,
soften the CR spectrum and reduce the CR acceleration efficiency.

\acknowledgements 
The author was supported by National Research Foundation of Korea 
through grant 2010-0016425.


\end{document}